\documentclass[preprint,12pt]{elsarticle}
\usepackage{graphicx}
\usepackage{booktabs}
\usepackage{subcaption}
\usepackage{comment}
\usepackage{amssymb}
\usepackage{float}
\usepackage{lipsum}
\usepackage{ulem}
\usepackage{caption}
\usepackage{supertabular}

\usepackage{orcidlink}

\usepackage{longtable}
\usepackage{adjustbox}
\usepackage{rotating}
\usepackage{tikz}
\usepackage{geometry}

\usetikzlibrary{shapes.geometric, arrows, positioning}
\usepackage{tabularx}
\usepackage{natbib}
\usepackage{pdflscape}

\UseRawInputEncoding

\usepackage{amsmath}
\usepackage{hyperref}

\usepackage[T1]{fontenc}

\usepackage{makecell}
\usepackage{booktabs}
\usepackage{multirow}

\usepackage{color}

\begin{document}
\biboptions{numbers,sort&compress}
\begin{frontmatter}
\title{Interpretation of the binned SNe Ia Master Sample data via a scalar quintessence component: phantom transition?}

\author[1,7]{Giovanni Montani\fnref{fn1}}
\fntext[fn1]{giovanni.montani@enea.it}

\author[1]{Iolanda Navone\fnref{fn4}}
\fntext[fn4]{navone.1867377@studenti.uniroma1.it (Corresponding author)}

\author[2,3,4]{Maria Giovanna Dainotti\fnref{fn2}}
\fntext[fn2]{maria.dainotti@nao.ac.jp}

\author[6]{Shigehiro Nagataki\fnref{fn4}}\fntext[fn4]{shigehiro.nagataki@riken.jp}

\affiliation[1]{Physics Department, Sapienza University of Rome, P.le A. Moro 5, 00185 Roma, Italy}

\affiliation[7]{ENEA, Nuclear Department, C.R. Frascati, Via E. Fermi 45, 00044 Frascati, Italy}

\affiliation[2]{Division of Science, National Astronomical Observatory of Japan, 2-21-1 Osawa, Mitaka 181-8588, Tokyo, Japan}
\affiliation[3]{Graduate University for Advanced Studies (SOKENDAI), 2-21-1 Osawa, Mitaka, Tokyo 181-8588, Japan}
\affiliation[4]{Space Science Institute, 4765 Walnut St Ste B, Boulder, CO 80301, USA}

\affiliation[6]{Astrophysical Big Bang Laboratory (ABBL), RIKEN Pioneering Research Institute (PRI), 2-1 Hirosawa, Wako, Saitama 351-0198, Japan}
\affiliation[7]{RIKEN Center for Interdisciplinary Theoretical and Mathematical Sciences (iTHEMS), 2-1 Hirosawa, Wako, Saitama 351-0198, Japan}
\affiliation[8]{RIKEN-Berkeley Center (RBC), RIKEN iTHEMS, 2-1 Hirosawa, Wako, Saitama 351-0198, Japan}

\begin{abstract}
We study a modified cosmological scenario for the late Universe, involving an evolutionary dark energy model associated with the dynamics of a self-interacting scalar field in a potential-dominated regime. Through the analogy with a fluid energy-momentum tensor, we introduce a viscous contribution to the scalar dynamics, accounting for effective non-equilibrium behaviour  of the self-interacting scalar cluster. The resulting picture is that of an intrinsic quintessence contribution which, due to the bulk viscosity, admits an effective equation of state parameter that can also take values below $-1$. Within this framework, we set up the diagnostic tool of the so-called ``effective running Hubble constant'', which allows us to trace possible deviations from a standard $\Lambda$CDM model. We then compare this theoretical function with binned data from the Master Sample of Supernovae Ia~\cite{Dainotti:2025qxz}, constructed assuming a $\Lambda$CDM model in the MCMC procedure performed in each bin. We show that the self-interacting scalar field corresponding to the best fit satisfies a slow-rolling condition, since the kinetic energy remains small compared to the potential contribution throughout the redshift interval. The key finding is that, when limiting the model to specific regions of the parameter space and fitting it to the data, the transition only occurs at redshifts significantly lower than the value $z \sim 0.5$ identified by the DESI Collaboration~\cite{DESI:2025zgx, DESI:2024mwx}. Furthermore, for the parameter values ensuring the best fit, no quintessence-to-phantom transition occurs (i.e., the effective equation of state parameter remains below $-1$ across the whole redshift domain). In other words, Supernova data alone provide no indication of a change in the nature of the dark energy.
\end{abstract}

\end{frontmatter}

\section{Introduction}

One of the most pressing issues in contemporary cosmology is the persistent Hubble tension, characterized by a significant ($>4\sigma$) discrepancy between local measurements of the Hubble constant $H_0$ \citep{SH0ES} and those inferred from early-Universe observations \citep{Planck2018}. To address this tension, the community has pursued diverse independent measurement strategies probing a wide redshift range (see for example \citep{Dainotti2023, Dainotti2022a, Dainotti2024PDU....4401428D, Mukherjee2025, Dainotti2024Galax..12....4D, Dainotti2023a, Dainotti2022b, Bargiacchi2025, Adil2024, Abbott2017,Kalita2026, Faucher2026, Gadbail2026, Zhan2026}). Despite this multiplicity of approaches, the tension remains unresolved, suggesting either unaccounted systematic effects or the need for new physics beyond the standard $\Lambda$CDM framework.

In this context, a surprising step forward was made in the last few years with the analysis of Baryonic Acoustic Oscillations (BAO) data on large scales in the Universe, as provided by the DESI Collaboration \citep{DESI:2025zgx, DESI:2024mwx}.

The relevance of this achievement lies, first of all, in the determination of the Chevallier-Polarski-Linder (CPL) model (\cite{CPL1,CPL2}) as the best fit to the data when including Planck satellite observations {\citep{Planck2018}, BAOs and SNe Ia, i.e., it is statistically preferred over the standard $\Lambda$CDM model according to its Deviance Information Criterion (DIC) (see Tab. 3 in \cite{DESI:2025fii} for the analysis' results using DESI+CMB data combined with PantheonPlus, Union3, DESY5, with a favorable $\Delta$DIC of $-6.8, -13.5, -17.2$ respectively). 
The same preference can also be seen when computing the Bayesian evidence (see Tab. 4 in \cite{DESI:2025fii}; the only exception is the DESI+CMB+ PantheonPlus dataset, which instead shows a light preference for $\Lambda$CDM). Although conclusions are not yet settled \cite{ong2026bayesianviewdesidr2, efstathiou2025evolvingdarkenergysupernovae}, this evidence strongly suggests a possible evolutionary nature of dark energy, rather than a constant vacuum contribution (for recent analyses, see \citep{Capozziello_2026, Chaudhary_2026, Chaudhary_2026_2, chaudhary2026evidenceevolvingdarkenergy, Chaudhary_2025, reboucas2026modelingnonlinearscalesdynamical, ishak2025persistentchallengelambdacdmthrone, park2025updatedobservationalconstraintsphicdm, alfano2025cosmicdistancedualitydesi, 2024JCAP...12..007C, Wolf_2024, Wolf_2025, Wolf_2026}). In \cite{Cort_s_2024} it was argued that the DESI findings depend on the choice of prior in the 
CPL parameter space, questioning the robustness of the claimed deviation from $\Lambda$CDM. Calderon et al.~\cite{2024JCAP...10..048C}, 
on the other hand, recovered consistent hints of evolving dark energy 
through a non-parametric reconstruction independent of the CPL 
parametrization, lending further support to the DESI conclusions.

A second important outcome of the DESI Collaboration analysis is the detection of the so-called \textit{phantom transition}, i.e., a redshift value that separates a later quintessence phase of dark energy \citep{Tsujikawa_2013} from an earlier phantom one \citep{Caldwell_2002, PhysRevD.68.023509}. This result has renewed  interest for the Quintom Dark Energy scenario \citep{c1,c2,c3}.

The CPL model is a valuable but purely phenomenological representation of the physical properties of dark energy. The presence of a phantom phase for $z > 0.5$ has puzzling implications when addressing which physical mechanisms govern late-time Universe dynamics (it also implies a "PhantomX Coincidence"~\cite{Cort_s_2024}: for the transition to happen, the dark energy must have reached its maximum density exactly in the redshift range best constrained by the data). 

Possible evidence for a decaying effective (or running) Hubble parameter \citep{schiavone2024, dainotti2021, Fazzari:2025mww} has already been outlined in binned analyses of Type Ia supernova data, as discussed in \citep{dainotti2021, dainotti2022, Dainotti:2025qxz, Krishnan:2020vaf, colgain_Dainotti}.

As clarified in \citep{Fazzari:2025mww}, this decaying behaviour  of the effective Hubble parameter, when interpreted in the presence of phantom dark energy, is mainly associated with a continuous phantom regime (in contrast with the phantom transition highlighted by DESI results). In fact, a decaying effective Hubble parameter appears to be more consistent with possible resolutions of the Hubble tension \citep{divalentino-Hubbletension, CosmoVerseNetwork:2025alb, Montani:2025rcy, Montani_carlevaro_dainotti, Montani_deangelis_dainotti, Montani:2024pou, efstathiou2021, elizalde_odintsov}.

At present, the most accurate representation of the decaying profile observed in binned data relies on a power-law form (see \cite{Dainotti:2025qxz} for a recent detailed analysis), which is motivated by the possibility of a redshift evolution in the Standard Candle relation parameters (see \cite{Dainotti_2013, desimone2024doubletcosmologicalmodelschallenge}).

However, this power-law behaviour  has also been reproduced via well-identified physical mechanisms: in \citep{schiavone2023}, through a metric $f(R)$ gravity theory, and in \citep{Montani:2025rcy}, through a dark energy-dark matter interaction scenario (see also \cite{Dainotti:2026logvspow} for a comparison between power-law and logarithmic behaviour s, the latter derived in \cite{leclair2025quantumvacuumenergyorigin} based on a Casimir-like effect).

In the present study, we combine two physically motivated ingredients: (i) dark energy can be described by a viable quintessence model, represented by the dynamics of a classical self-interacting scalar field in a slow-roll regime, i.e., when the potential energy dominates over the kinetic term \citep{Montani_carlevaro_dainotti}; (ii) the phantom-like behaviour  of the dark energy equation of state is interpreted as an effective phenomenon (since true phantom matter violates the strong energy condition \citep{PhysRevD.68.023509}), arising from the combination of quintessence with deviations from thermodynamic equilibrium, modeled via a perturbative bulk viscosity term \citep{Montani_carlevaro_dainotti}.

Specifically, we consider a classical self-interacting scalar field subject to an additional condition that controls the relative contributions of kinetic and potential energy, while also incorporating non-equilibrium effects. This way, we obtain a closed dynamical system with three unknowns: the scale factor of the Universe (equivalently, the Hubble parameter), the scalar field, and its potential as a function of time. The functional form of the potential governing the scalar dynamics is then derived \textit{a posteriori}.

The key feature of our theoretical framework is that the dark energy component has an intrinsic quintessence-like equation-of-state parameter which, due to the presence of bulk viscosity, effectively evolves into a phantom regime beyond a certain redshift. In this way, we are able to reproduce a phantom transition whose origin is grounded in a consistent physical mechanism.

The paper is organized as follows. In Section~2, we introduce the physical framework underlying the model, discussing the correspondence between a self-interacting scalar field and an effective viscous fluid description. In Section~3, we develop the cosmological dynamics of the model in a flat Friedmann-Lemaitre-Robertson-Walker Universe and define the effective running Hubble constant used as the main diagnostic tool. Section~4 presents the Supernovae Ia Master Sample dataset together with the statistical methodology and MCMC analysis adopted for the fits. In Section~5, we discuss the obtained results, focusing on the effective equation of state, the possible phantom transition, and the consistency of the perturbative regime. Finally, Section~6 summarizes the main conclusions and outlines possible future developments involving additional cosmological probes such as DESI BAO data.

\section{Physics background}

It is well known \cite{Weinberg:1972kfs} that, in a spacetime with metric tensor $g_{\mu\nu}$ 
($\mu, \nu = 0,1,2,3$), the energy-momentum tensor of a classical self-interacting scalar field $\phi$, 
described by a potential term $V(\phi)$, takes the following form:

\begin{equation}
	T^{(\phi)}_{\mu\nu} = 
	\partial_{\mu}\phi\partial_{\nu}\phi - \mathcal{L}g_{\mu\nu}
	\, ,
	\label{eq:emtensor}
\end{equation}

where the Lagrangian density $\mathcal{L}$ reads as

\begin{equation}
	\mathcal{L}\equiv 
	\frac{1}{2}\partial_{\rho}\phi\partial^{\rho}\phi - V(\phi)
	\, . 
	\label{eq:lagrangian}
\end{equation}

The energy-momentum tensor~(\ref{eq:emtensor}) can be mapped into that of a 
perfect fluid (having energy density $\varepsilon$, pressure $p$, and $4$-velocity 
$u_{\mu}$), i.e.,

\begin{equation}
	T_{\mu\nu}^{(pf)} \equiv 
	\left( \varepsilon + p \right) u_{\mu}u_{\nu} - p g_{\mu\nu}
	\, ,
	\label{eq:perfectfluid}
\end{equation}

provided the following formal identities hold \cite{Montani:2009hju}: 

\begin{equation}
	\varepsilon \equiv 
	\frac{1}{2}\partial_{\rho}\phi \partial^{\rho}\phi + V \, ,\quad
	p \equiv \mathcal{L} 
	\, ,\quad
	u_{\mu}\equiv \frac{\partial_{\mu}\phi}{\sqrt{\partial_{\rho}\phi\partial^{\rho}\phi}}
	\, . 
	\label{eq:identities}
\end{equation}

Here, we extend the correspondence above to the case in which a bulk viscosity contribution is associated with fluid dynamics, meaning that a weak non-equilibrium process is affecting the scalar field physics.

The case of bulk viscosity characterizing a relativistic fluid is well modeled \cite{Montani:2009hju, Weinberg:1972kfs, Belinskii1975, Belinskii1977, Belinskii1979, Disconzi_2015, Montani_2017,Carlevaro:2005dt} by a negative pressure 
contribution $\bar{p}\equiv - \xi \nabla_{\mu}u^{\mu}$, where $\nabla_{\mu}$ denotes 
covariant differentiation and $\xi$ is the bulk (second) viscosity coefficient, which typically depends 
on thermodynamical parameters. 

Thus, according to the identities~(\ref{eq:identities}), we add to the energy-momentum tensor~(\ref{eq:emtensor}) of a 
self-interacting scalar field the following negative pressure contribution:

\begin{equation}
	\bar{p}\equiv  
	- \xi \nabla_{\mu}\left( 
	\frac{\partial^{\mu}\phi}{\sqrt{\partial_{\rho}\phi\partial^{\rho}\phi}}\right)
	\, . 
   \label{eq:bulkpressure}
\end{equation}

We now attempt to assign a precise physical meaning to this ansatz. In cosmology, a classical 
massive free scalar field can be naturally interpreted as a Bose condensate, a bosonic state with an extremely high occupation number. This situation arises at the end of inflation, when the inflaton reaches the true vacuum \cite{2008cosm.book.....W}, or in the late Universe, when 
axion physics is addressed \cite{Kolb:1990vq,Peacock:1998eq}. 
Hence, we can infer that a classical self-interacting scalar field represents a bosonic 
condensate cluster of particles: this is clearly the case as long as 
the self-interaction term is sufficiently weak. However, 
in what follows, we will be interested in a situation in which the potential term dominates over the kinetic contribution. In this limit, the concept of ``particles'' is difficult to recover at the microscopic level, as occurs in Quantum Chromodynamics at low energy scales \cite{Peskin:1995ev}.

Nonetheless, the idea of a self-interacting mode characterized by a very high occupation number, i.e.,  
overlapping a large number of interacting momentum values, appears to be a plausible representation of a self-interacting scalar field in a cosmological context. 

In this theoretical scenario, the negative pressure contribution introduced in Eq.~(\ref{eq:bulkpressure}) is a phenomenological approach 
to describe a weak degree of non-equilibrium physics in the self-interacting boson cluster. Such non-equilibrium phenomenology could arise from 
the competition between the Universe expansion and the strength of the self-interaction coupling, as well as from specific collisional effects within the self-interaction itself.

On the basis of these theoretical conjectures and the associated technical tools, we now develop the cosmological implementation of Eq.~(\ref{eq:bulkpressure}), restricting our attention to the 
case in which the self-interacting scalar field corresponds to a quintessence model \cite{Tsujikawa_2013}.

\section{Cosmological Dynamics}

We now consider a flat isotropic 
Universe, in accordance with the Planck Satellite observations as discussed in 
\cite{efstathiou_planck}, which is described by the line element

\begin{equation}
	ds^2 = dt^2 - a(t)^2 dl^2
	\, , 
	\label{eq:lineelement}
\end{equation}

where $t$ denotes the synchronous time 
(we use $c=1$ units), $dl^2$ stands for the Euclidean line element, and 
$a(t)$ denotes the cosmic scale factor of the Universe, which regulates the 
expansion of every physical distance. 

We consider the late Universe as 
characterized by the presence of a 
matter contribution (dark and baryonic) and a self-interacting scalar 
field, i.e., the Friedmann equation 
reads as follows \cite{Montani:2009hju}:

\begin{equation}
	H^2 \equiv \left(\frac{\dot{a}}{a}\right)^2 = \frac{\chi}{3}
	\left( \varepsilon_m + 
	\frac{1}{2}\dot{\phi}^2 + V(\phi)\right)
	\, .
	\label{eq:friedmann}
\end{equation}

Above, $\chi$ is the Einstein constant, 
the dot denotes differentiation with 
respect to the synchronous time, and 
$\varepsilon_m$ is the matter 
energy density, whose evolution is 
regulated by the following relations:

\begin{equation}
	\dot{\varepsilon}_m + 3H\varepsilon_m = 0 \, \Rightarrow \, 
	\varepsilon_m = \varepsilon_m^0/a^3 \equiv \varepsilon_m^0(1+z)^3
	\, , 
	\label{eq:matterdensity}
\end{equation}

where $\varepsilon_m^0$ denotes the 
present-day value of the matter 
energy density (we set, by convention, the present-day value of the scale factor equal to unity) and we have introduced the redshift variable $z$. 

The energy density of the scalar field, 
$\varepsilon_{\phi}\equiv \dot{\phi}^2/2 + V$, obeys the equation corresponding to 
the $0$-component of the vanishing 
divergence of the energy-momentum tensor~(\ref{eq:emtensor}) with the additional 
negative pressure of Eq.~(\ref{eq:bulkpressure}), i.e.: 

\begin{equation}
	\dot{\varepsilon}_{\phi} = 
	-3H\left( \varepsilon_{\phi} + p_{\phi} + \bar{p}_{\phi}\right) = -3H\dot{\phi}^2 + 9H^2\xi
	\, , 
	\label{eq:scalarfieldenergy}
\end{equation}

where we have introduced the subscript $\phi$ also for the two (positive and negative) pressure contributions. We 
stress that, for a purely time-dependent scalar field $\phi(t)$, as 
Universe homogeneity prescribes, the 
$4$-velocity associated with it 
is clearly a comoving one. 

Since we are interested in investigating the dynamics of the scalar field when it is able to reproduce an equation of state near $p_{\phi}\simeq -\varepsilon_{\phi}$, corresponding to a small deviation from a $\Lambda$CDM model, we 
impose on our system the following condition: 

\begin{equation}
	\dot{\phi}^2 = \beta V(\phi)
	\, ,\quad \beta = \mathrm{const.} \ll 1
	\, .
	\label{eq:betacondition}
\end{equation}

Imposing this condition implies, 
for dynamical consistency, that the potential term 
$V(t)\equiv V(\phi(t))$ becomes one of the problem's unknowns: the form of the self-interaction potential $V(\phi)$ is 
determined \textit{a posteriori}, once the function 
$\phi(t)$ is inverted (which therefore needs to be monotonic). 

It is straightforward to verify that the 
scalar field's intrinsic (i.e., in the absence of bulk viscosity) equation of state parameter takes the simple form

\begin{equation}
	w_{\phi}
	\equiv \frac{p_{\phi}}{\varepsilon_{\phi}} = 
	\frac{\beta/2 - 1}{\beta/2 + 1}
	\, .
	\label{eq:eos}
\end{equation}

Thus, we see that our classical self-interacting scalar field corresponds macroscopically to 
a quintessence cosmological fluid. 
In fact, the equation of state above 
still describes dark energy if we 
relax the condition in Eq.~(\ref{eq:betacondition}) 
to the weaker constraint $\beta < 1$.

Now, in agreement with the analysis 
in \cite{navone2025}, we consider the following natural assumption 
for the bulk viscosity coefficient: 

\begin{equation}
	\xi = \xi(\varepsilon_{\phi}) = \bar{\xi}\varepsilon_{\phi} 
	\, ,\quad \bar{\xi}=\mathrm{const.}
	\, . 
	\label{eq:bulkcoeff}
\end{equation}

It is easy to see that Eq.~(\ref{eq:scalarfieldenergy}) can be associated with the 
following effective scalar field 
equation, once Eq.~(\ref{eq:betacondition}) is taken into account:

\begin{equation}
	\ddot{\phi} + 3H\left( 
	1 - \frac{\beta + 2}{2\beta}
	3\bar{\xi}H\right)\dot{\phi} + \frac{dV}{d\phi} = 0
	\, . 
	\label{eq:scalarfieldequation}
\end{equation}

Now, introducing the new time 
variable $x\equiv \ln(1+z)$, such that 
$\dot{(\cdots)} \equiv -H\,d(\cdots)/dx$, Eq.~(\ref{eq:scalarfieldenergy}) can be restated via 
Eq.~(\ref{eq:betacondition}) as follows: 

\begin{equation}
	\frac{dV}{dx} = 3 
	\left( \frac{2\beta}{2+\beta} 
	- 3H\bar{\xi}\right) V
	\, .
	\label{eq:Vevolution}
\end{equation}

Introducing the dimensionless critical parameters

\begin{equation}
	\Omega_m \equiv \frac{\chi\varepsilon_m}{3H_0^2}\, ,\quad 
	\Omega_\Lambda\equiv \left( 
	1 + \frac{\beta}{2}\right) 
	\frac{\chi V}{3H_0^2}
	\, , 
	\label{eq:omegaparams}
\end{equation}

where $H_0\equiv H(x=0)$ is the 
Hubble constant, the Friedmann 
equation~(\ref{eq:friedmann}) can be restated as 
follows:

\begin{equation}
	E^2 = \Omega_m^0 e^{3x} 
	+ \Omega_\Lambda
	\, , 
	\label{eq:Efriedmann}
\end{equation}

where $E\equiv H/H_0$ is the Universe 
expansion rate and $\Omega_m^0 \equiv 
\Omega_m(x=0)$. 

Furthermore, Eq.~(\ref{eq:Vevolution}) 
takes the dimensionless form:

\begin{equation}
	\frac{d\Omega_\Lambda}{dx} = 
	3\left( \bar{\beta} - \gamma E\right)\Omega_\Lambda
	\, , \quad \Omega_\Lambda(x=0) = 1 - \Omega_m^0 \, ,  
	\label{eq:omegalambdaevolution}
\end{equation}

where 
$\gamma \equiv 3\bar{\xi}H_0$ 
and $\bar{\beta} \equiv 2\beta/(2+\beta)$. 
It is important to stress that the 
bulk viscosity effect, given our perturbative approach, is a physical contribution 
only as long as the following condition holds:

\begin{equation}
	\frac{|\bar{p}_{\phi}|}{|p_{\phi}|} = \frac{3\bar{\xi}H_0 E}{|w_{\phi}|} \ll 1
	\, .
	\label{eq:perturbativity}
\end{equation}

In this respect, the effective equation 
of state parameter of our scalar system 
reads as follows:

\begin{equation}
	w_{\phi}^{\mathrm{eff}} \equiv 
	w_{\phi} - 3\bar{\xi}H_0 E
	\, . 
	\label{eq:weff}
\end{equation}

Since $\beta \ll 1$ and $E$ is 
expected to be an increasing function of $x$ (mainly due to the matter contribution), two considerations are 
in order with respect to Eqs.~(\ref{eq:perturbativity}) and~(\ref{eq:weff}):
(i)~the interval of $x$ in which the 
bulk viscosity remains perturbative is 
larger when the value of 
the parameter $\bar{\xi}$ is smaller;
(ii)~if $w_{\phi}^{\mathrm{eff}}(x=0) > -1$, 
i.e., the scalar system dynamics corresponds today to a quintessence fluid, 
sooner or later an instant $x_{cross}$ 
will exist at which 
a phantom transition takes place, 
in accordance with the DESI Collaboration 
analysis \cite{DESI:2025zgx, DESI:2024mwx}.

\subsection{\label{sec:level2}The effective running Hubble constant}

Previous studies (\cite{Fazzari:2025mww, Montani:2025rcy,PhysRevD.103.103509,krishnan2022h0universalflrwdiagnostic,schiavone2024}) have shown that the \textit{running Hubble constant}, defined as

\begin{equation}
\mathcal{H}_{0}(z) \equiv H_0\frac{E(z)}{E_{\Lambda CDM}(z)}=\frac{H(z)}{\sqrt{ \Omega^0_{m} (1+z)^{3} + 1 - \Omega^0_{m} }} \,,
\end{equation}

is a useful object for discerning the nature of dark energy, since it is tailor-made to highlight the differences in behaviour between dynamical dark energy models and the $\Lambda$CDM model. In the above equation, in fact, $E_{\Lambda \text{CDM}}$ denotes the Universe expansion rate according to the $\Lambda$CDM model in late Universe dynamics. If the Universe follows the $\Lambda$CDM model, this function is expected to reduce to the constant value $\mathcal{H}_{0}(z) \equiv H_0$; however, the appearance of a non-constant $\mathcal{H}_{0}(z)$ reflects the spread in the different experimental values of $H_0$ when observed in different redshift intervals.

For a dynamical dark energy model as the one considered in this paper, using Eq. \eqref{eq:Efriedmann} we obtain:

\begin{equation}
\mathcal{H}_{0}(z) =H_0\frac{\sqrt{ \Omega^0_{m} (1+z)^{3} + \Omega_{de}(z) }}{\sqrt{ \Omega^0_{m} (1+z)^{3} + 1 - \Omega^0_{m} }} \,,
\label{eq:Hrunning}
\end{equation}

This is the theoretical observable that will be used to test our model.

\section{\label{sec:analysis}Data and methodology}

The Master binned Sample of SNe Ia originally compiled in \cite{Dainotti:2025qxz} is the dataset chosen to analyze the model. It collects 3714 SNe Ia drawn from the Dark Energy Survey \citep{DESy5}, Pantheon+ \citep{Scolnic_2022, Brout_2022}, Pantheon \citep{scolnic2018} and the Joint Lightcurve Analysis \citep{JLA}, with no duplicates (see Tab. \ref{tab:sn_sample}). Apart from combining the largest number of SNe Ia so far in a unified dataset, the Master Sample has also been constructed with a novel statistical analysis that takes into account the most appropriate likelihoods following \cite{DAINOTTI202430}, which are not necessarily the Gaussian ones (see also \cite{Dainotti_2023, Bargiacchi_2023}).
\begin{table}[h!]
\centering
\caption*{\textbf{Master binned Sample}}
\vspace{-2mm} 
\begin{tabular}{l c}
\hline
Survey & Number of SNe Ia\\
\hline
\addlinespace[5pt]
DES & 1829 \\
Pantheon+ & 1208 \\
Pantheon & 181 \\
JLA & 496 \\
\addlinespace[5pt]
\hline
\end{tabular}
\vspace{1mm}
\caption{Sample of Type Ia Supernovae collected from various surveys to construct the Master Sample.}
\label{tab:sn_sample}
\end{table}
The Master Sample has been divided into $20$ bins of equal population, covering the range from $z=0.00122$ to $z=2.3$. The choice of $20$ bins aims for a balance between limiting the statistical uncertainties for each data point - by providing enough supernovae inside each bin - and keeping a sufficient number of data points for accurate fitting. During the binning procedure, implemented in \cite{Dainotti:2025qxz}, it was also verified that the inferred best-fit parameters and their uncertainties remain stable when varying the binning scheme (e.g. changing the number of bins, implementing the moving window method, or using the $\log(z)$ division which takes into account the volumetric segregation). This also indicates that residual inter-bin correlations do not significantly affect the inferred uncertainties and can be neglected.

The observed distance modulus $\mu_{obs}$ as a function of the apparent magnitude can be taken from the average in each redshift interval and compared with the theoretical expectation that follows from a standard $\Lambda$CDM framework. This theoretical luminosity distance is denoted as $d_L(z)$: 

\begin{equation}
d_L(z) = (1+z)\frac{c}{H_0}\int_0^z\frac{\text{d}z'}{\sqrt{\Omega_m(1+z')^3+\Omega_\Lambda}} .
\end{equation}

The observed luminosity distance is instead extracted from the distance modulus:

\begin{equation}
\mu_{\mathrm{obs}} = m_B^\ast - M_B + \alpha\,x_1 - \beta\,c + \Delta_M + \Delta_B ,
\end{equation}

\begin{equation}
\mu_{\mathrm{th}}(z) \;=\; 5\log_{10}\!\left(\frac{d_L(z)}{\mathrm{Mpc}}\right) + 25 \, ,
\end{equation}

\noindent where: $m_B^\ast$ is the observed rest-frame peak magnitude in the $B$ band; $\alpha$ and $\beta$ are nuisance parameters that quantify the correlations between luminosity and, respectively, the light-curve shape parameter ($x_1$) and the color parameter ($c$); $\Delta_M$ is a distance correction accounting for the empirical tendency of SNe Ia in more massive galaxies to appear brighter; $\Delta_B$ corrects bias derived from survey simulations; $M_B$ is the absolute B-band magnitude of a fiducial SN Ia
with $x_1 = 0$ and $c = 0$ (degenerate with $H_0$). \noindent To remove the degeneracy, $M_B$ has been calibrated using a reference value of $H_0=70$ km/s/Mpc, since it does not influence the descending trend of $\mathcal{H}_0(z)$. 

Carrying out a Markov Chain Monte Carlo (MCMC) procedure in each bin \footnote{This code has been developed by the authors from an original set-up written by Elisa Fazzari}, estimates of $H_0$ and $\Omega_m^0$ have been extracted for each redshift interval, and assumed to be statistically independent from neighbouring bins (for the reason explained above). The prior for $H_0$ has been taken to be flat over the interval 
$[60,\,80]~\mathrm{km\,s^{-1}\,Mpc^{-1}}$, 
while the prior for $\Omega_{m}^0$ is Gaussian, $\mathcal{N}\bigl(\mu=0.322,\sigma=0.025\bigr)$.
The values for the Gaussian prior have been derived at 5 $\sigma$ after performing a fit of the whole sample using the $\Lambda$CDM model. This is how the dataset was constructed, containing therefore $20$ values of $\mathcal{H}_0(z)$ and $\Omega_m^0$ measurements.

By construction, the dataset therefore approximates the cosmological dynamics as following the $\Lambda$CDM model for small redshift intervals. When using this dataset to test new cosmological models, rigour may be at risk for models that deviate significantly from the $\Lambda$CDM predictions (we are assuming this is not the case for our model, as long as the slow-rolling condition is preserved). 

By numerically integrating the system of equations \eqref{eq:Efriedmann} and \eqref{eq:omegalambdaevolution}, we compared the theoretical running Hubble constant predicted by the model in Eq. \eqref{eq:Hrunning} with the experimental values of $\mathcal{H}_0(z)$, i.e., the trend for $H_0$ revealed in the data.

We explored the model's behaviour for different values of $\bar{\beta}$, spanning orders of magnitude from $10^{-4}$ to its maximum possible value ($\bar{\beta}=2/3$, corresponding to $\beta=1/3$). For each value, an MCMC procedure was performed to fit the model to the Master binned sample, constraining the free parameters $H_0,\Omega_m^0,\xi$ and evaluating the quality of fit. 

The priors on $H_0$ and $\Omega_m^0$ (see Tab. \ref{tab:prior}) have been chosen in order for the procedure to be consistent with the analysis in \cite{Fazzari:2025mww} and \cite{navone2025}, while the prior for $\bar{\xi}$ is motivated by our need to keep the bulk viscosity contribution small enough to be perturbative, as will soon be shown. Four parallel chains have been used for each fit, and the Gelman-Rubin criterion $R-1<0.01$ has been implemented to establish chain convergence \citep{gelman_rubin}. Finally, statistical results and plots have been produced using \texttt{GetDist} \citep{getdist}.

\begin{table}[ht!]
\centering
\begin{tabular}{l l}
\hline
\addlinespace[2pt]
\textbf{Parameter} & \textbf{Prior} \\
\hline
\addlinespace[5pt]
$\Omega_m^0$ & $U[0.01, 0.99]$ \\
$H_0$ (km/s/Mpc) & $U[60, 80]$ \\
$\bar{\xi}$ & $U[0, 0.01]$ \\
\addlinespace[5pt]
\hline
\end{tabular}
\caption{Prior distributions for the model parameters used in the MCMC procedure.}
\label{tab:prior}
\end{table}

\section{Results}

Since our goal was to study whether the quintessence to phantom transition is statistically favoured (for the chosen model with the chosen dataset) we examined the redshift at which the crossing occurs ($z_{cross}$) as a function of the parameter $\bar{\beta}$. The results, plotted in Fig. $\ref{fig:crossing}$, show that even if $z_{cross}$ increases for increasing values of $\bar{\beta}$, it never reaches the redshift $z_{cross}\simeq 0.5$ predicted by the DESI Collaboration. Furthermore, in Fig. \ref{fig:chi} there seems to be an increasing trend in the $\chi^2$ for high values of $\bar{\beta}$, while the minimum $\chi^2$ is around $\bar{\beta}\simeq 0.1$, an order of magnitude that does not result in a quintessence to phantom transition at any redshift.

\begin{figure}[H]
    \centering
    \begin{subfigure}[b]{0.49\textwidth}
        \centering
        \includegraphics[width=\textwidth]{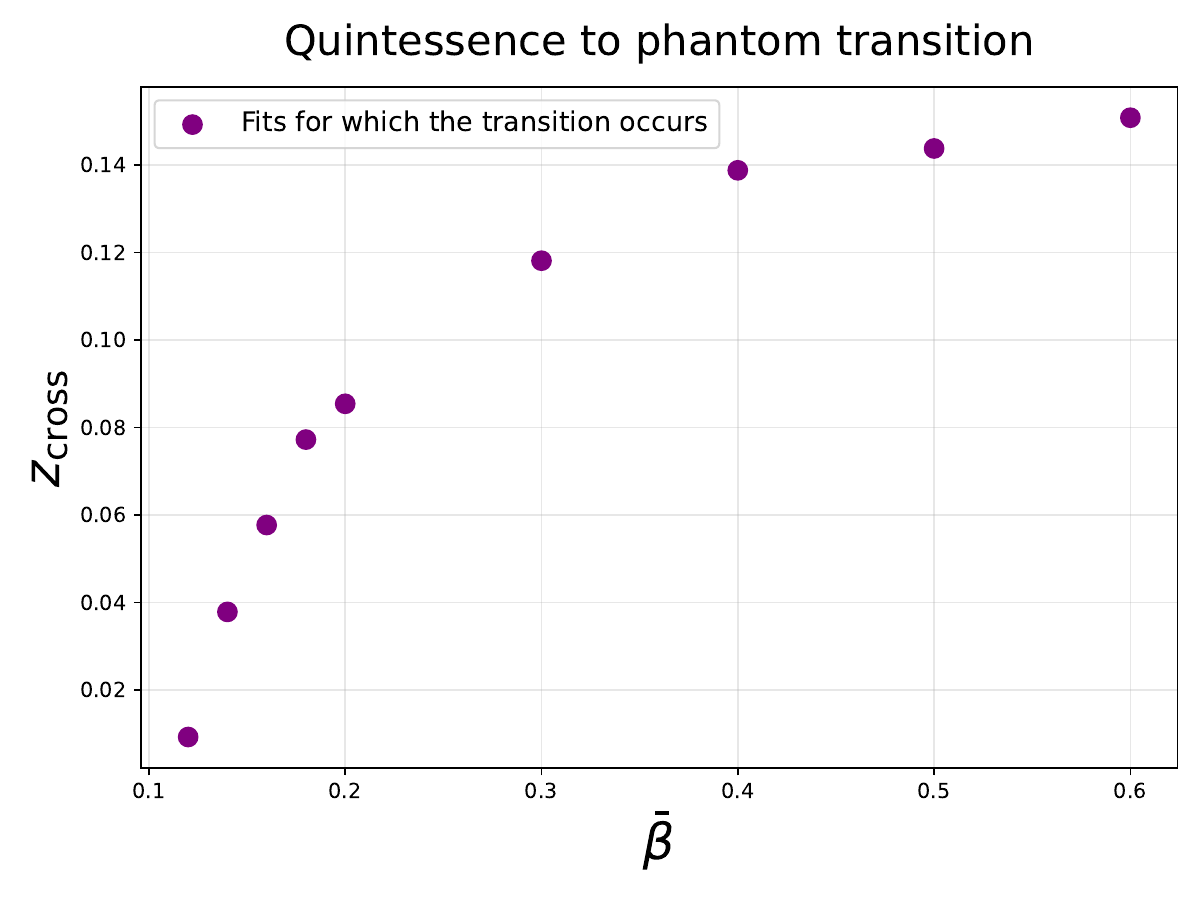}
        \caption{Redshift at which the quintessence to phantom transition occurs as a function of the chosen fixed values for $\bar{\beta}$. The transition only occurs for $\bar{\beta}>0.1$.}
        \label{fig:crossing}
    \end{subfigure}
    \hfill
    \begin{subfigure}[b]{0.49\textwidth}
        \centering
        \includegraphics[width=\textwidth]{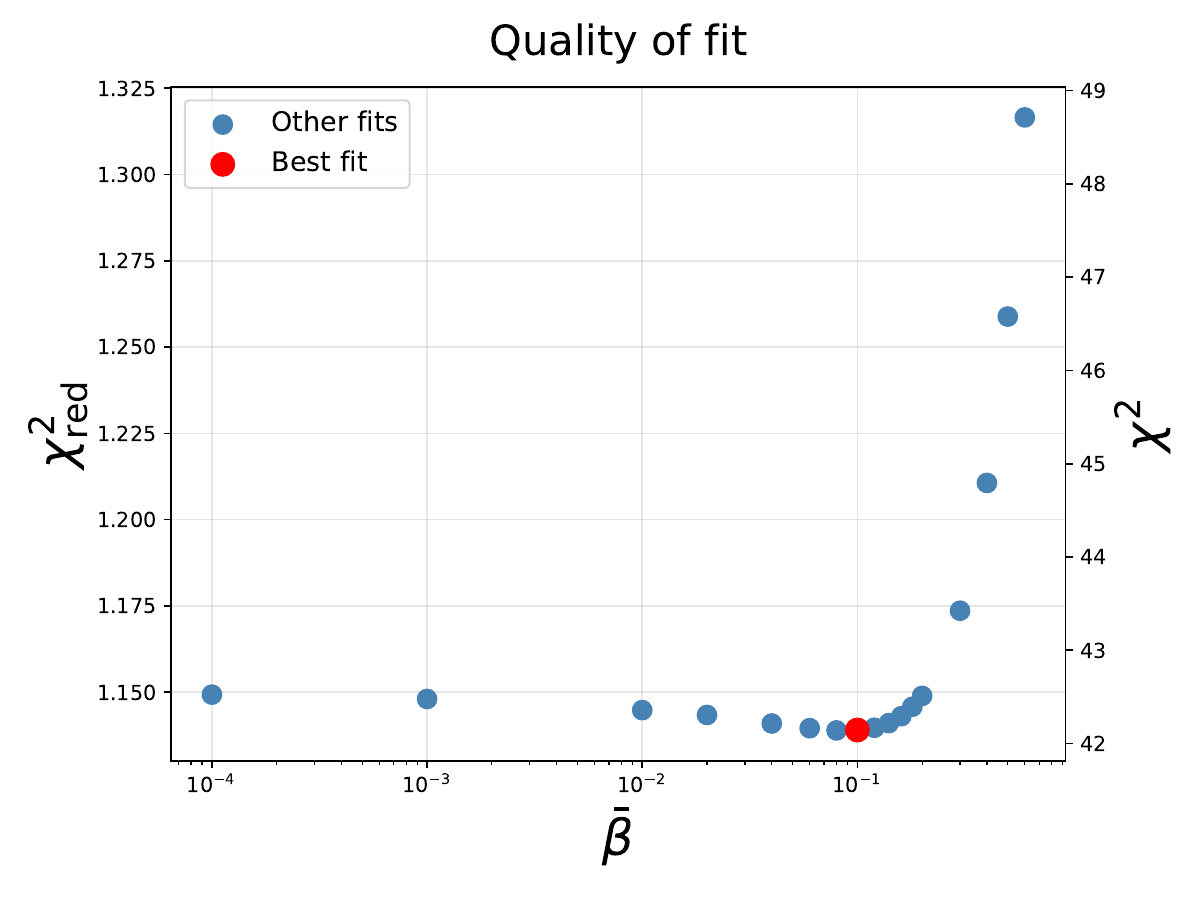}
        \caption{Reduced $\chi^2$ corresponding to different fits with the chosen fixed values for $\bar{\beta}$; $\bar{\beta}=0.1$ corresponds to the best-fit.}
        \label{fig:chi}
    \end{subfigure}
    \caption{Comparison between (on the left) the values of $\bar{\beta}=2\beta/(2+\beta)$ for which the cosmological model fitted to the Master Sample shows a quintessence to phantom transition, and (on the right) the trend of the quality of fits for the same $\bar{\beta}$. It can be observed that the lowest values of the $\chi^2$ are reached when the transition occurs at very late times (z<0.1), or not at all.}
    \label{fig:both}
\end{figure}

The following plots in Fig. \ref{fig:fit} and \ref{fig:weff} show the results of the different fits and the corresponding effective equation of state parameters, for selected values of $\bar{\beta}$ (for a qualitatively  similar $w(z)$ behaviour corresponding to  a model-agnostic analysis, see Fig. 6 in \cite{2025JCAP...02..021A}); the free parameters for the fit with the lowest $\chi^2$, corresponding to $\bar{\beta}=0.1$, are listed in Tab. \ref{tab:fit_results} (with posteriors shown in Fig. \ref{fig:post}). Finally, the function of the potential $V(z)$ has been derived from the integration of $\Omega_\Lambda$, according to Eq. \eqref{eq:identities}, and is plotted in Fig. \ref{fig:V}. 

\begin{figure}[H]
    \centering
    \includegraphics[width=0.8\textwidth]{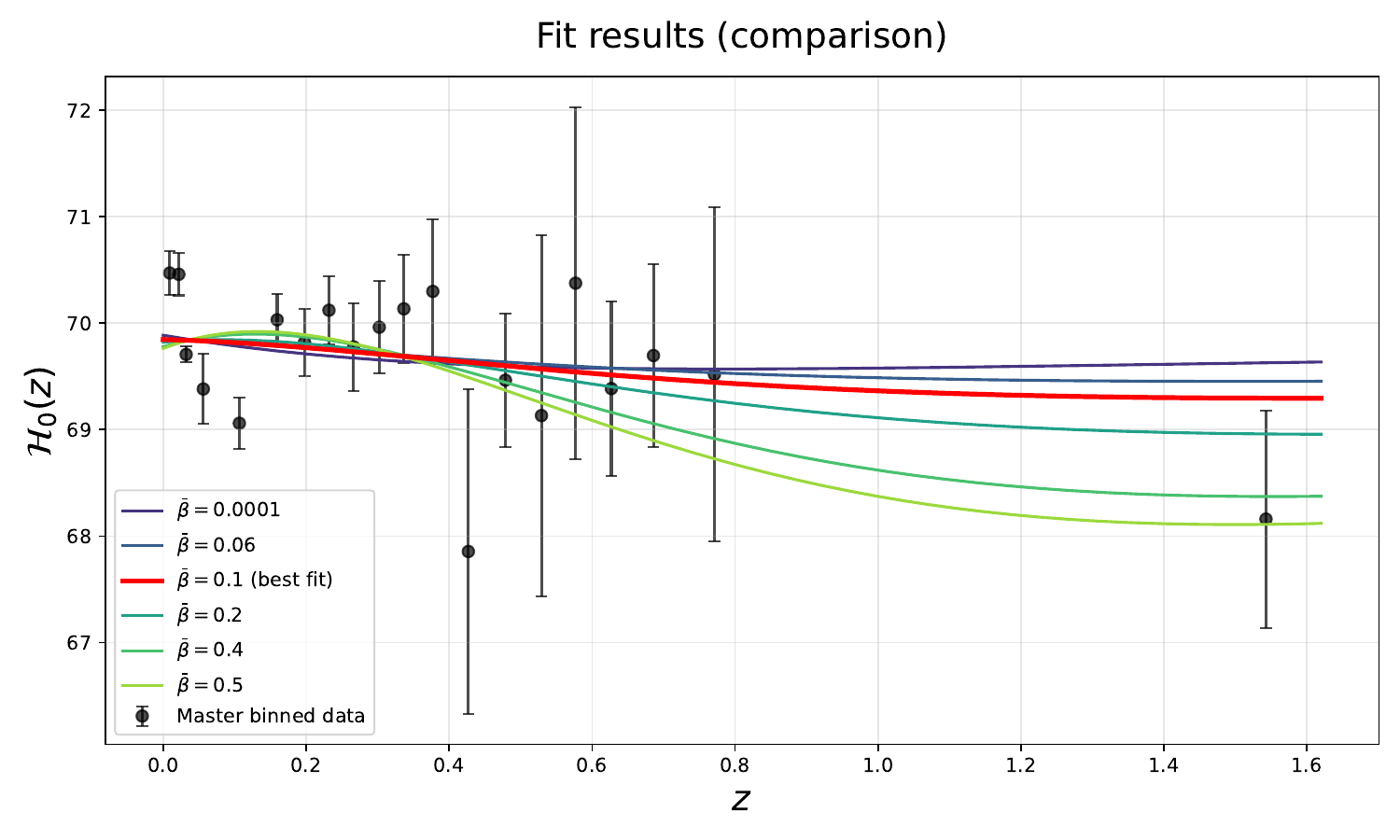}
    \caption{Fit of the effective running Hubble constant of the cosmological model using the Master Sample binned data, numerically integrating Eq.\eqref{eq:omegalambdaevolution} for different fixed values of $\bar{\beta}$ and performing an MCMC procedure with the prior in Tab. \ref{tab:prior}. The fit with the lowest $\chi^2$ is shown in red, and the corresponding fitted parameters can be found in Tab. \ref{tab:fit_results}.}
    \label{fig:fit}
\end{figure}

\begin{figure}[H]
    \centering
    \includegraphics[width=0.8\textwidth]{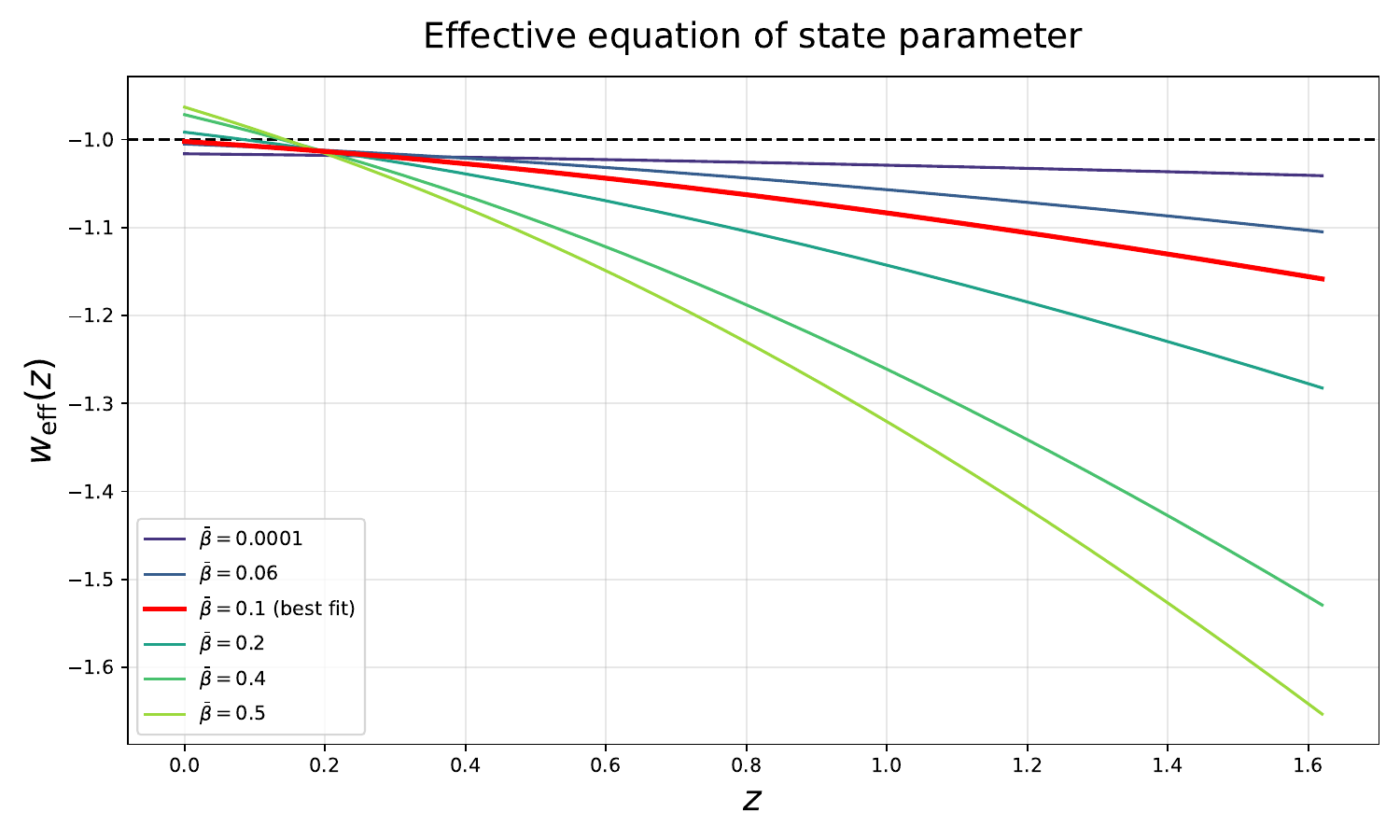}
    \caption{Trend of the effective equation of state parameter in Eq. \eqref{eq:weff} (using also Eq. \eqref{eq:eos}) after the fit, for different values of  $\bar{\beta}=2\beta/(2+\beta)$. The quintessence to phantom transition can be seen for $\bar{\beta}<0.1$, at the redshift shown in Fig. \ref{fig:crossing}. The function corresponding to the fit with the lowest $\chi^2$ is shown in red.}
    \label{fig:weff}
\end{figure}

\begin{figure}[H]
    \centering
    \begin{subfigure}{0.8\textwidth}
        \centering
        \includegraphics[width=\textwidth]{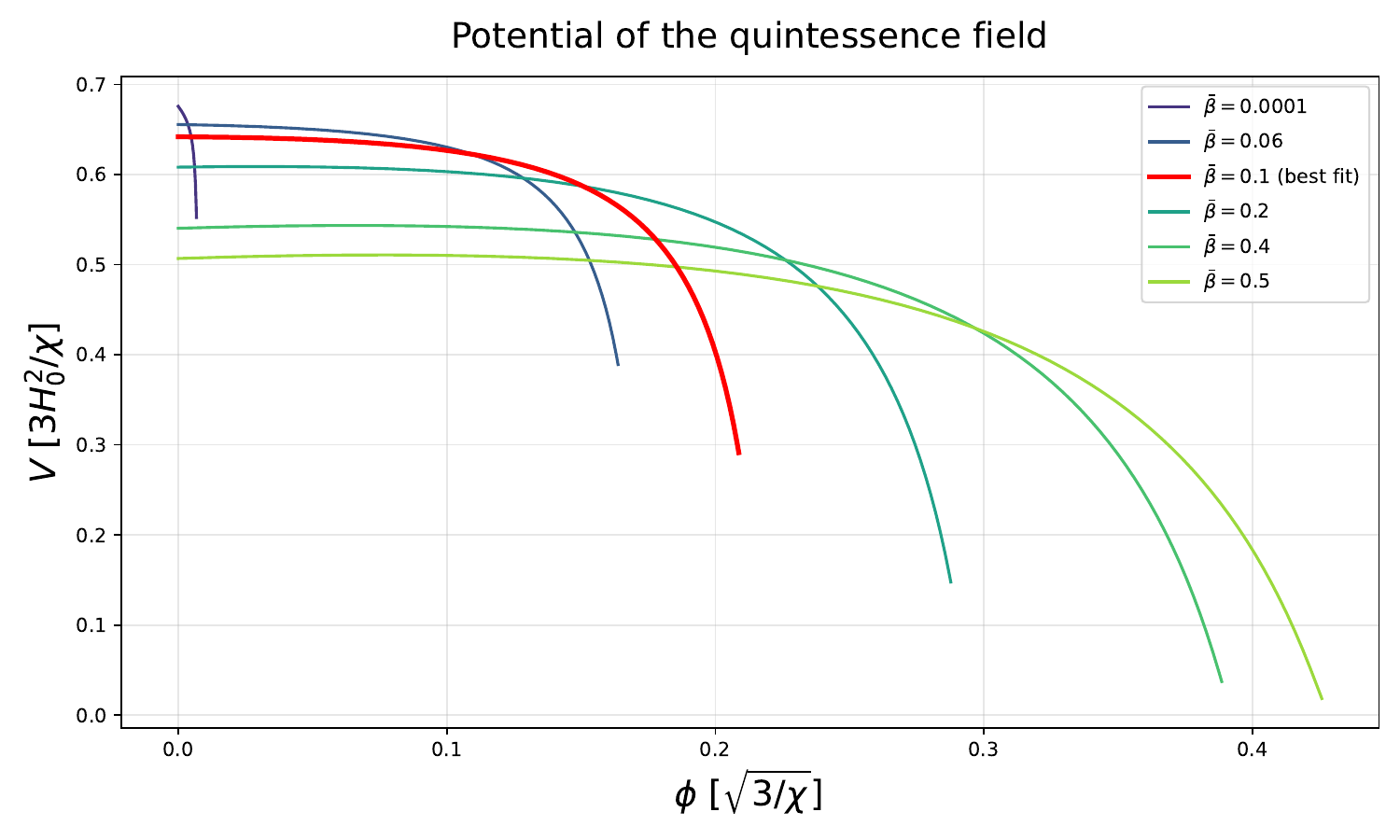}
    \end{subfigure}
    \hfill
    \begin{subfigure}{0.8\textwidth}
        \centering
        \includegraphics[width=\textwidth]{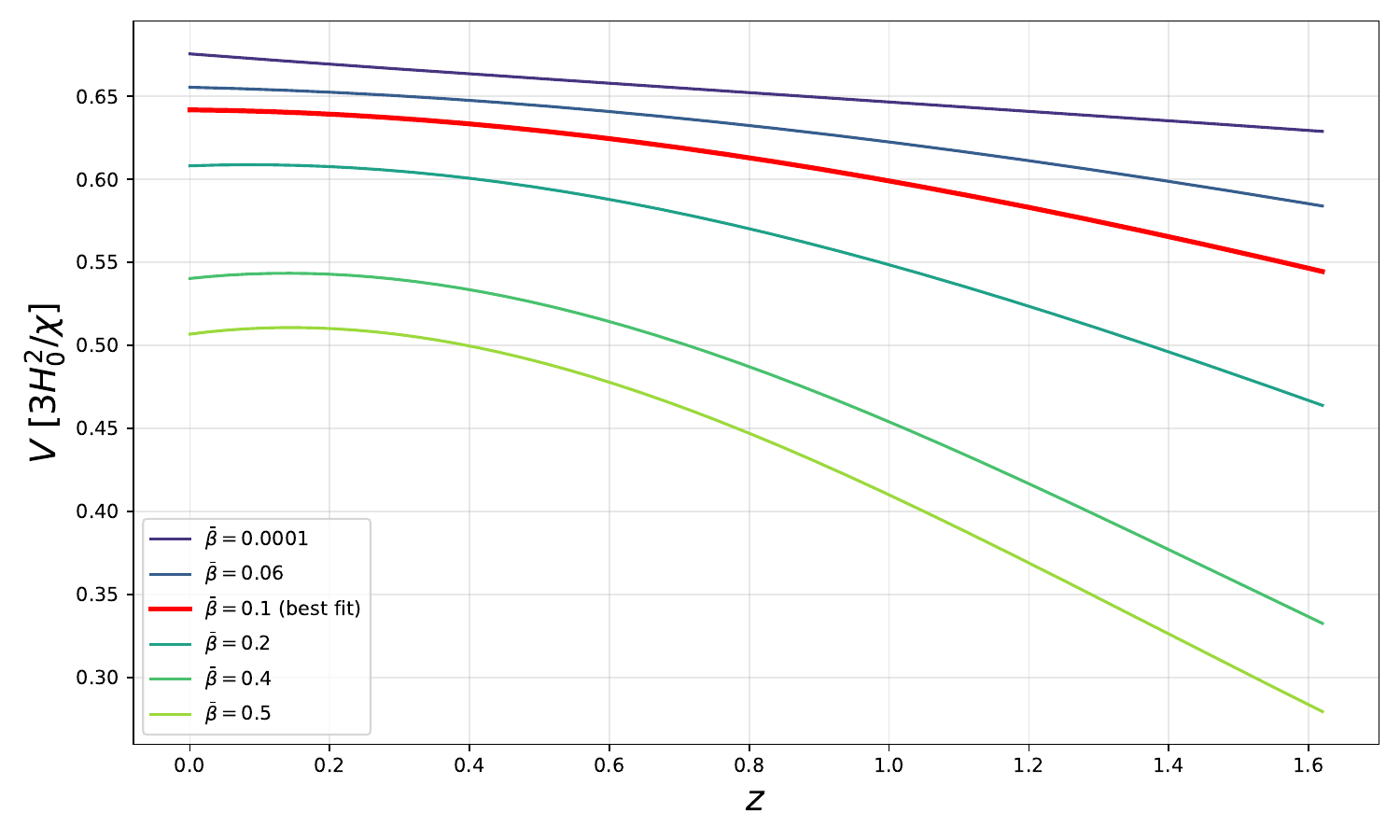}
    \end{subfigure}
    \caption{Trend of the potential numerically integrated using Eq. \eqref{eq:scalarfieldequation} and Eq. \eqref{eq:betacondition}, for different values of  $\bar{\beta}=2\beta/(2+\beta)$ and as a function of the quintessence field (top) and the redshift (bottom). The integration stops in $\phi(z_{max}\simeq 1.6)$ when it reaches the redshift interval of the Master Sample: for the lowest values of $\bar{\beta}$, $\phi$ evolves more slowly with the redshift, so the range covered by the integration is shorter. The function corresponding to the fit with the lowest $\chi^2$ is shown in red. In general, near $z\simeq 0$ the slow rolling condition appears to be satisfied.}
    \label{fig:V}
\end{figure}

\begin{table}[H]
\centering
\renewcommand{\arraystretch}{1.4}
\begin{tabular}{lc}
\toprule
\textbf{Parameter} & \textbf{Value} \\
\midrule
$H_0$ [km/s/Mpc]      & $69.847 \pm 0.077$ \\
$\Omega_m^0$            & $0.3244 \pm 0.0055$ \\
$\bar{\beta}$         & $0.1$ (fixed) \\
$\bar{\xi}$           & $0.000487 \pm 0.000060$ \\
\midrule
$\chi^2_\mathrm{red}$ & $1.14$ \\
\bottomrule
\end{tabular}
\caption{Best-fit parameters using the Master Sample of Supernovae Ia.}
\label{tab:fit_results}
\end{table}

\begin{figure}[H]
    \centering
    \includegraphics[width=0.7\textwidth]{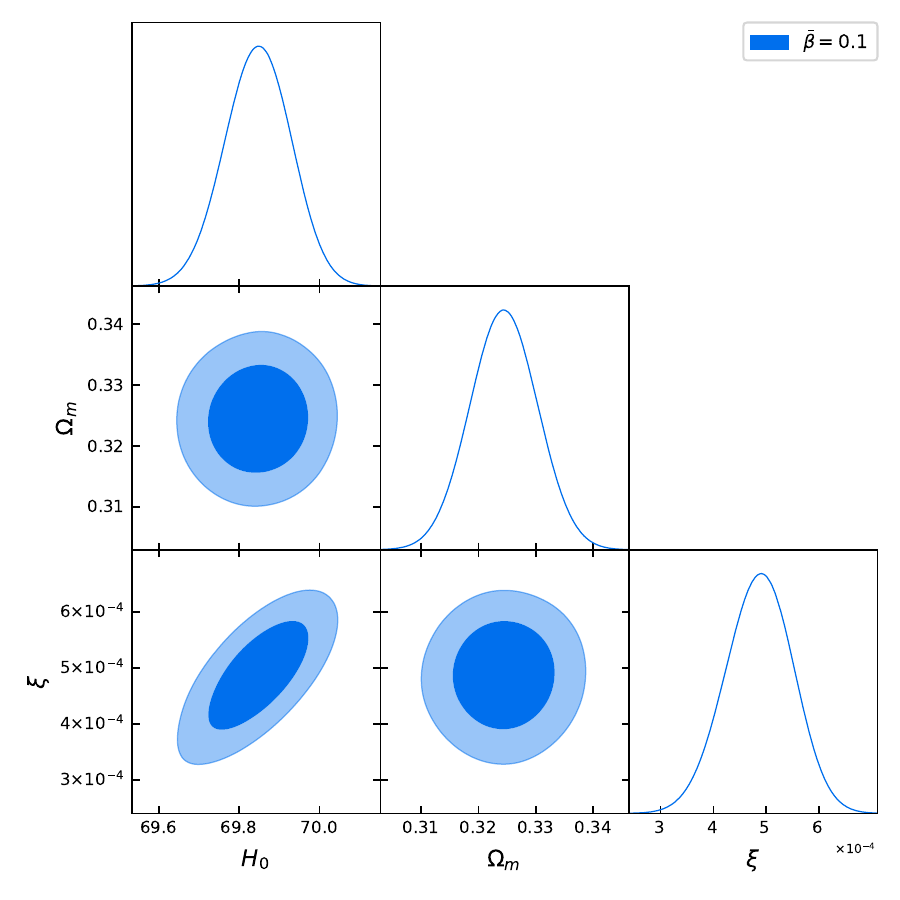}
    \caption{Posteriors for the parameters in Tab. \ref{tab:fit_results}, corresponding to the fit of the cosmological model with $\bar{\beta}$ using the Master Sample. The priors used for the MCMC procedure are listed in Tab. \ref{tab:prior}.}
    \label{fig:post}
\end{figure}

Lastly, from Fig. \ref{fig:pert} one can infer that for $\bar{\beta}<0.2$ the fitted bulk viscosity coefficient $\xi$ stays in an order of magnitude lower than $10^{-3}$ and the perturbative condition (see Eq. \eqref{eq:perturbativity}) consequently holds in the redshift interval concerning the dataset. For higher values of $\bar{\beta}$ (and therefore $\xi$) the model loses internal consistency.

\begin{figure}[H]
    \centering
    \includegraphics[width=0.8\textwidth]{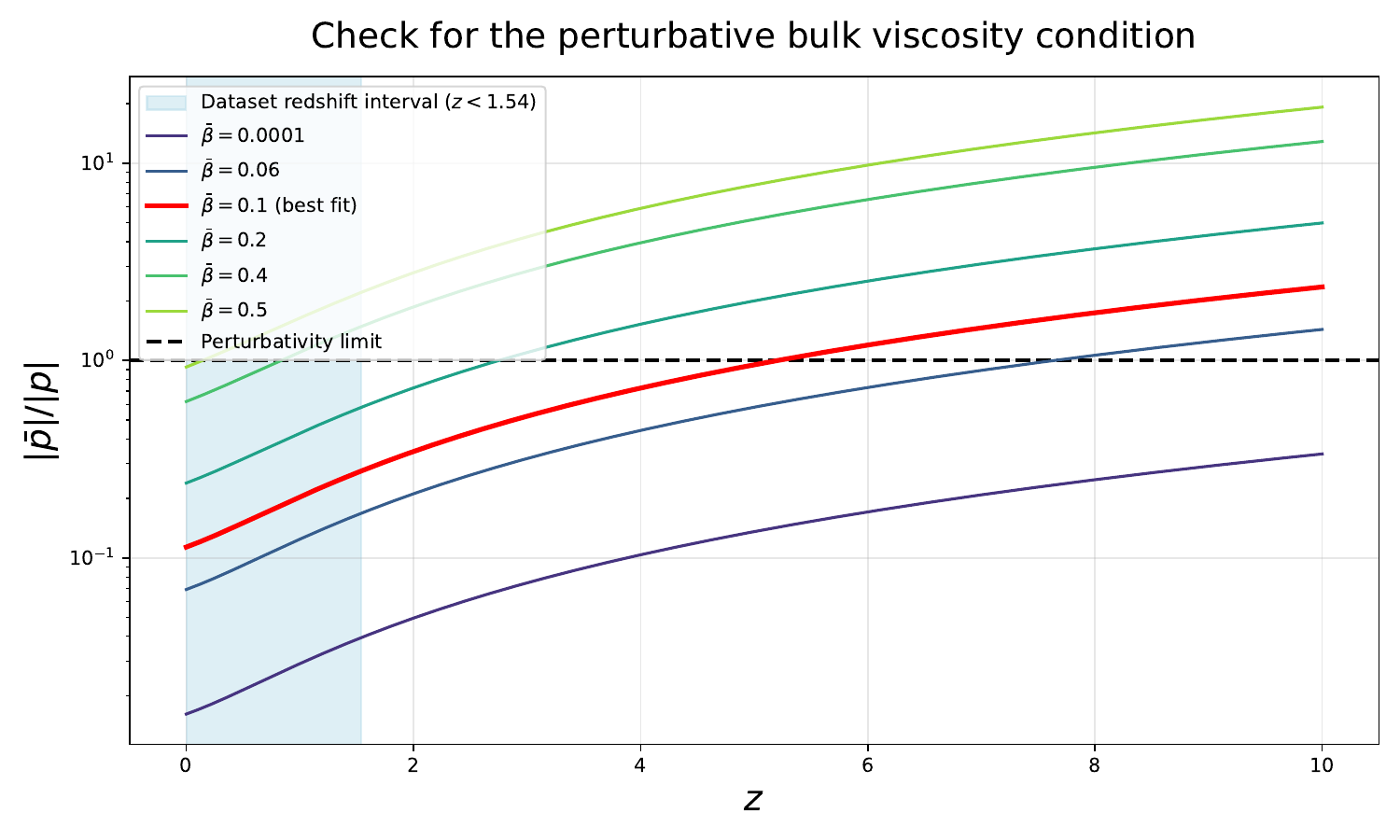}
    \caption{Ratio of the bulk viscosity contribution to the total equilibrium pressure (see Eq. \eqref{eq:perturbativity}) for the fits plotted in Fig. \ref{fig:fit}. This plot shows at which redshift the model crosses the perturbativity limit, and therefore loses the assumption of perturbative bulk viscosity that the model was built with. For the redshift interval concerning the binned Master Sample dataset, fits with values of $\bar{\beta}$ of order of magnitude $10^{-1}$ or lower can be considered fully reliable.}
    \label{fig:pert}
\end{figure}

\section{Conclusions}

We constructed a cosmological scenario describing the Universe's dark energy contribution via a classical self-interacting scalar field in a potential-dominated regime. The analogy with a perfect fluid energy-momentum tensor allowed us to attribute a bulk viscosity phenomenology to the dark energy component, arising from a deviation from thermodynamic equilibrium in the boson cluster associated with the self-interacting particles (in the limit of high occupation numbers).

In the proposed picture, the dark energy component is also associated with an effective equation of state parameter that, depending on the redshift variable, can drop below $-1$, thus potentially realizing a phantom transition in agreement with the observations by the DESI Collaboration~\cite{DESI:2025zgx, DESI:2024mwx}. 

Using the Hubble parameter obtained for our revised late Universe cosmology, we constructed the effective running Hubble constant as a diagnostic tool to interpret the binned data of the SNe Ia Master Sample \cite{Dainotti:2025qxz}. The fitting procedure favoured values around $\bar{\beta} \simeq \beta \sim 0.1$, with the corresponding bulk viscosity parameter $\bar{\xi}$ turning out to be of order $5 \times 10^{-4}$.

These results have two important consequences. From a theoretical point of view, a value of $\beta \simeq 0.1$ indicates that the scalar field dynamics undergoes a slow-rolling evolution, since by Eq.~(\ref{eq:betacondition}) the kinetic energy is a small fraction of the potential term. Furthermore, from a phenomenological point of view, we arrive at the conclusion that the values of the model's free parameters associated with the fitting procedure are unable to ensure a phantom transition in the redshift regime where it is observed by the DESI Collaboration.

Thus, we can conclude that even in a model that allows for a passage from a quintessence nature to a phantom one for dark energy, the information coming from SNe Ia -- see also the previous analyses in Refs.~\cite{Fazzari:2025mww, navone2025} -- does not support a change in the character of the dark energy component of the Universe. This conclusion motivates further investigations of the present dynamical proposal when other sources, and in particular BAO-DESI data, are included in the analysis.

\section{Acknowledgements}
M.G.D. and S.N. acknowledge the support of the JSPS Grant-in-Aid for Scientific Research (KAKENHI) (A), Grant Number JP25H00675. S.N. acknowledges the support of KAKENHI (B), Grant Number 23K25874. I. N. acknowledges the support of the Sokendai Asian Spring School. M.G.D. thanks the support of NAOJ accomodation facilities and DoS at NAOJ for hosting I.N. supervised by M.G.D. during her two weeks internship.

\bibliographystyle{elsarticle-num}
\bibliography{biblio2}

\end{document}